# Communicating Two States in Perovskite Revealed by Time-Resolved Photoluminescence Spectroscopy

Yanwen Chen[#,1], Tianmeng Wang[#,1], Zhipeng Li[1,2], Huanbin Li[1,3], Tao Ye[3], Christian Wetzel[4], Hanying Li[3,*], Su-Fei Shi[1,5,*]

[1] The Department of Chemical and Biological Engineering, Rensselaer Polytechnic Institute, Troy, NY 12180

[2] School of Chemistry and Chemical Engineering, Shanghai Jiao Tong University, Shanghai, 200240, China

[3] MOE Key Laboratory of Macromolecule Synthesis and Functionalization, State Key Laboratory of Silicon Materials, Department of Polymer Science and Engineering, Zhejiang University, Hangzhou, 310027, P. R. China.

[4] The Department of Physics, Applied Physics and Astronomy, Rensselaer Polytechnic Institute, Troy, NY 12180

[5] The Department of Electrical, Computer and Systems Engineering, Rensselaer Polytechnic Institute, Troy, NY 12180

[#] These authors contributed equally to this work

*Corresponding author: hanying_li@zju.edu.cn, shis2@rpi.edu

**Organic-inorganic perovskite as a promising candidate for solar energy harvesting has attracted immense interest for its low-cost preparation and extremely high quantum efficiency. However, the fundamental understanding of the photophysics in perovskite remains elusive. In this work, we have revealed two distinct states in $MAPbI_3$ thin films at low temperature through time-resolved photoluminescence spectroscopy (TRPL). In particular, we observe a photo-induced carrier injection from the high energy (HE) state to the low energy (LE) state which has a longer lifetime. The strong interaction between the two states, evidenced by the injection kinetics, can be sensitively controlled through the excitation power. Understanding the interacting two-states not only sheds light on the long PL lifetime in perovskite but also helps to understand the different behavior of perovskite in response to different excitation power. Further efforts in modifying the low energy state could significantly improve the quantum efficiency and lead to novel application in optoelectronics based on perovskite.**

## 1. Introduction

Hybrid organic-inorganic perovskites such as $CH_3NH_3PbI_3$ ($MAPbI_3$) have attracted intense research interest worldwide recently for their promising application in solar energy harvesting[1–6]. $MAPbI_3$ obtained from low-cost solution based processing has enabled high-efficiency solar cell devices[3,7–11]. With the material development and device optimization, record high efficiencies exceeding 22% have been demonstrated[9,12–15]. Perovskites also possess superior optical properties such as large optical gain and low lasing threshold[16–20]. However, the fundamental mechanism of carrier excitation and recombination remains elusive. In particular, solar cell application requires a long carrier lifetime to enable a long diffusion length (considering that the mobility of the $MAPbI_3$ is reasonably good, on the order of ~ 100 $cm^2V^{-1}s^{-1}$)[21–23], while lasing requires a recombination rate high enough to outcompete nonradiative channels. These two requirements seem to contradict each other and cannot be satisfied in the same material system.

We have performed continuous wave and time-resolved photoluminescence (TRPL) spectroscopy of $MAPbI_3$ at low temperature. In this way, we first revealed two well-separated states, a high energy (HE) and a low energy (LE) state, in PL spectroscopy with continuous wave excitation. Then in time-resolved spectroscopy, we analyzed the dynamics between the two states in the time domain. We find their strong coupling and a direct communication between these two states. We find that optical excitation populates the HE state with carriers which then are efficiently injected into the separate LE state. The rate of this injection we find to be a sensitive function of the optical excitation power. We believe that this strong coupling of the two communicating states should be responsible for the long PL lifetime in perovskites under low optical excitation power. In particular, we see evidence that communication to the LE state of a longer lifetime for temporary storage. Upon high excitation power, however, the LE state exhibits a saturation behavior and the response of the perovskite is found to be dominated by the HE state.

## 2. Results and Discussion

### 2.1 TRPL Data and Discussion

We prepared the $MAPbI_3$ thin film using the spin-coating method and the polycrystalline film was deposited on a silicon wafer with 300 nm thermal oxide[24,25]. To minimize sample degradation, the film was prepared under Argon gas in a glove box and quickly transferred to the vacuum ($< 10^{-6}$

Torr) of an optical cryostat for optical characterization. PL spectroscopy was performed using a confocal microscope setup with a spatial resolution of ~ 2 µm. TRPL was performed using the Time-Correlated Single Photon Counting (TCSPC) technique[26].

We first characterized the film in PL at room temperature and 77 K under 2.33 eV ($\lambda$ = 532 nm) continuous wave (CW) laser excitation with an excitation power of 1 µW (Fig. 1a). At room temperature, we find a single peak C at 1.61 eV. This finding is distinct from the observation of a double peak PL at room temperature that was attributed to bulk and surface recombination respectively in previous studies[27,28]. Yet, the 77 K spectrum is distinctively different from that at room temperature. The spectrum now exhibits one HE peak A at 1.65 eV and one LE peak B at 1.58 eV at 77 K. This splitting is not limited to certain areas of the film. Instead, we find it uniformly present across the entire film (see SI Section S2 for spectra at various positions).

As we increase the excitation power from 1 µW to 10 µW (Fig. 1c), the ratio of the PL intensity ($I_A/I_B$) increases from 0.91 to 1.56. Once the excitation power exceeds 2 µW, peak A dominates. Both peak intensities $I$ can be fitted with a power law of the excitation power $P$ as $I \sim P^\alpha$ (Fig. 1d). We find $\alpha = 1.13$ for peak A and $\alpha = 1.02$ for peak B. The super linear dependence of the PL from peak A gives rise to the fast growing PL from peak A over peak B. The room temperature PL can also be fitted with a power law with $\alpha = 1.13$ (see SI Section S2.), consistent with the previous reports[29].

TRPL measurements under excitation fluence of 4.0 µJ/cm² at 2.61 eV ($\lambda$ = 475 nm) (frequency doubled Ti:Sapphire at a repetition frequency of 80 MHz) is shown in Fig. 1b. For peaks A (black, room temperature) and C (red, 77 K) we observe the typical instantaneous rise of the PL signal followed by a slow decay. Yet, peak B (blue, 77 K) behaves distinctively different in that the rise after the excitation pulse occurs asymptotically with a delay on the time scale of hundreds of picoseconds. A similar behavior had been previously shown in MAPbBr$_3$ single crystals[30].

To explore the rising delay of peak B in TRPL, we varied the excitation fluence (Fig. 2). The normalized signal over time of peaks A and B as a function of fluence are shown in a color contour plot in Fig. 2c and Fig. 2d, respectively. It is evident that peak A rises instantaneously after excitation, followed by an exponential decay. The peak position in time as indicated by the dashed line does not vary with increasing fluence as typical for a delocalized band state. However, peak B behaves distinctly different (Fig. 2d). With increasing excitation fluence, its rise time delay decreases and the delay disappears entirely for the highest fluence applied.[31] The difference can also be seen in the typical TPRL spectra at specific excitation fluence as shown in Fig. 2a and Fig. 2b, which corresponds to the specific line cuts in color plot Fig. 2c and Fig.2d, respectively. At the excitation fluence of 0.12 µJ/cm$^2$, the maximum of the TRPL signal from peak B occurs at 300 ps. This maximum TRPL position shifts close to time zero as the excitation fluence increases and remains at the same position as the excitation fluence exceeds 60 µJ/cm$^2$. For example, the TRPL of peak B at excitation fluence of 100 µJ/cm$^2$ exhibits the typical TPRL behavior with a sharp rise at time zero followed by an exponential decay (Fig. 2b).

**2.2 Two-level System Modelling**

The abovementioned observation can be explained with a two-level system, as schematically in Fig.3a. Here we are specifically interested in the quantification of the interaction and communication processes between the high energy (HE) and low energy (LE) states. The PL decay of peak A can be interpreted as the decay of the optically excited carriers in the high energy (HE) level, which is determined by the recombination rate $k_1$ (including both radiative and non-radiative channels) and the carrier injection rate $k_{12}$. $k_{12}$ depicts the carrier injection from the high energy (HE) level to the low energy (LE) level. We separate the $k_{12}$ process from other non-radiative processes for its contribution to the increased PL of peak B. This carrier injection, however, is sensitive to the available states in the LE level. As the excitation fluence increases, the maximum number of states in the perovskite will be occupied and the carrier injection pathway from HE to LE will be blocked, which leads to the decrease of $k_{12}$ and the disappearance of the rising feature (Fig. 2b and Fig. 2d). This interpretation also explains the fluence dependence of the PL spectra at 77 K (Fig. 1c) under the CW laser excitation: as the excitation power increases, the PL from peak A becomes the dominant one since the carrier injection channel is blocked.

A quantitative description is given by the following rate equations:
$$\frac{dn_1}{dt} = -k_1 n_1 - k_{12} n_1 \left(\frac{N_0 - n_2}{N_0}\right) \quad (1)$$

$$\frac{dn_2}{dt} = -k_2 n_2 + k_{12} n_1 \left(\frac{N_0 - n_2}{N_0}\right) \quad (2)$$

where $n_1$ and $n_2$ are optically excited carriers at the HE and LE state, respectively. $k_1$ is the decay rate for HE state, $k_2$ is the decay rate for the LE state, and $k_{12}$ is the injection rate of carriers from the HE to LE state. $N_0$ is the maximum number of states that can be occupied in the LE state. As shown in Fig. 3b, the TRPL data can be well fitted with this model. The obtained fitting parameters of $k_1$, $k_2$, and $k_{12}$ are plotted as a function of excitation fluence in Fig. 3c. It becomes apparent that the coupling rate $k_{12}$ is more than one order of magnitude larger than $k_1$ and $k_2$, respectively, suggesting a strong coupling between the LE and HE states.

It is this strong coupling that can explain the unusual decay of the TRPL spectra. In particular, by means of this strong coupling, Eq. (2) describes a delayed peaking at a time of hundreds of picoseconds away from time zero, qualitatively different from the typical TPRL which is peaked at time zero (within the resolution dictated by the response time of the avalanche photodiode (APD). To quantitatively compare the communication between the two states with the lifetime extracted from the exponential fitting of the TRPL data, we re-organize the Eq. (1) as $\frac{dn_1}{dt} = -(k_1+x)n_1$, in which x denotes the time average of the term $k_{12}\left(\frac{N_0-n_2}{N_0}\right)$ (i.e., $x = \frac{1}{\tau_1}\int_0^{\tau_1} k_{12}\left(\frac{N_0-n_2}{N_0}\right) dt$). We choose Eq. (1) instead of (2) for the relative similarity, i.e., both Eq. (1) and exponential decay describe a dynamic event with the peak value locates at the time zero. We therefore obtain a time scale $\tau_x$, through $1/x$, as a function of the excitation fluence, which is plotted along with the fluence dependence of $\tau_1$ and $\tau_2$ in Fig. 3d. It is clear that $\tau_2$, the lifetime of the LE state, is longer than the lifetime of the HE state, $\tau_1$. The difference is most drastic at the low excitation fluence. At the excitation fluence of 0.12 µJ/cm², $\tau_2$ (~ 9.5 ns) is more than three times as large as the $\tau_1$ (~ 3 ns). Considering the injection between the HE state and the LE state is particularly efficient under the low excitation fluence, the optically excited carrier can be transferred to the LE state with a longer lifetime.

## 2.3 Temperature Dependence

Previous studies have reported the observation of the two emission peaks in the low-temperature PL spectra of MAPbI$_3$ thin film[33–40]. While it is in general consensus that the HE energy state at 77K is attributed to the free exciton with a binding energy in the range of 20-60 meV[41,42], the nature of the LE state is elusive, with the possibility of a bound exciton[43], an exciton-receptor pair[37], and possible tetragonal phase domain in the orthorhombic phase[31,44]. One recent study suggests that the LE state is possibly due to an MA-disordered phase among the otherwise ordered orthorhombic phase[34,45,46]. To investigate the nature of the LE state, we performed the temperature dependent PL study in the range of 12K to 85 K (Fig. 4a). Since the temperature remains below 150 K the MAPbI$_3$ thin film remains in the orthorhombic phase (see SI Section S4). From 12 K to 85 K the PL peak shows a blue shift, consistent with the previous reports[33,34,36]. In parallel, the intensity ratio of HE peak and LE peak ($I_{HE}/I_{LE}$) decreased from 2.34 to 1.38. The increased LE PL at higher temperature suggests a thermal activation of the carrier injection from the HE to the LE state. More detailed temperature-dependent PL spectra can be found in SI Section S4. Fig. 4b shows the TRPL at 12 K and 85 K, and it is clear that the data at 85 K exhibits a slower rising, confirming the activation of the $k_{12}$[45,47]. The thermal activation behavior of the LE state is consistent with the MA-disordered domain picture. The kinetics extracted from TRPL, therefore, directly probe the interaction between the MA-ordered and MA-disordered domain. And such strong coupling may be achieved via electron-phonon interaction[23,38,48], energy transfer caused by dipole-dipole interaction or defect assisted scattering. More insight into the interaction mechanism requires further investigations.

## 3. Conclusion

In summary, we have identified two communicating states in the MAPbI$_3$ thin film through the low-temperature PL spectroscopy measurement. TRPL spectroscopy reveals a sensitive power dependence of the optically excited carrier injection from the HE state to the LE state, and the injection is particularly efficient at low power excitation. The LE state shows a thermal activation behavior, which might be attributed to the MA-disordered domain in the orthorhombic phase. The low-temperature TRPL spectroscopy directly probes the kinetics of the two states in MAPbI$_3$, which may enable the investigation of the modification effect on the LE state and help to improve our understanding of the optical properties of the perovskites.

## 4. Methods

### 4.1 Sample Preparation

The MAPbI$_3$ powder was synthesized and dissolved at a concentration of 1 mol/L in dimethylformamide (DMF). Microscope slides were washed sequentially with soap, de-ionized water, acetone, and isopropanol before they were finally treated under oxygen plasma for 20 minutes to remove the organic residues. The MAPbI$_3$ solution was spin-coated at 3000 rpm for 60 seconds, and the substrates were subsequently heated at 100 $^o$C on a hotplate in the glove box for 10 minutes to improve film quality.

**4.2 Optical Spectroscopy**

The steady-state photoluminescence (PL) and time-resolved photoluminescence (TRPL) spectroscopy measurements were performed with a home-built confocal microscope setup with either a CW or a femtosecond pulsed laser (repetition rate: 80 MHz). The excitation power of CW laser was typically maintained below 100 µW to prevent any sample degradation.

The TRPL measurement was performed by a Time-Correlated Single Photon Counting (TCSPC) module (PicoQuant TimeHarp-260) combined with an Avalanche Photo-Diode (MPD SPAD) through a spectrograph.

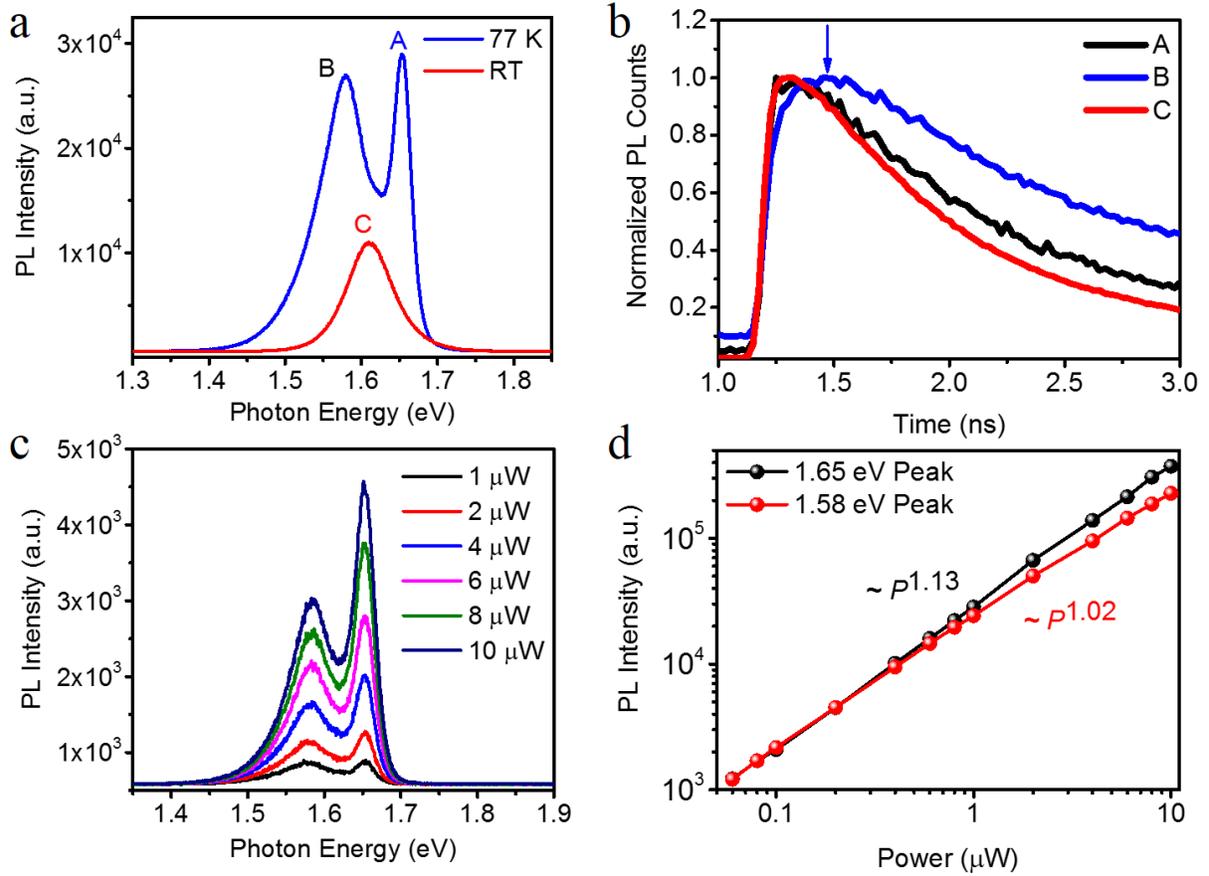

**Figure 1. Photoluminescence (PL) measurement of MAPbI$_3$ thin film**. (a) PL spectra at room temperature (red) and 77 K (blue) with the CW laser excitation centered at 2.33 eV ($\lambda$ = 532 nm). The excitation power is 1 μW, with a beam spot size of ~ 2 μm. (b) Normalized time-resolved PL (TRPL) at RT and 77 K with a pulsed (~ 120 fs pulse width) laser excitation centered at 2.61 eV ($\lambda$ = 475 nm). The excitation fluence is 4.0 μJ/cm$^2$. (c) Power dependence of PL spectra at 77 K. The LE peak amplitude is higher than that of the HE peak at low excitation power, while the HE peak becomes the dominant one with increasing power. (d) PL intensity of the HE (1.65 eV) peak and LE (1.58 eV) peak as a function of excitation power at 77 K.

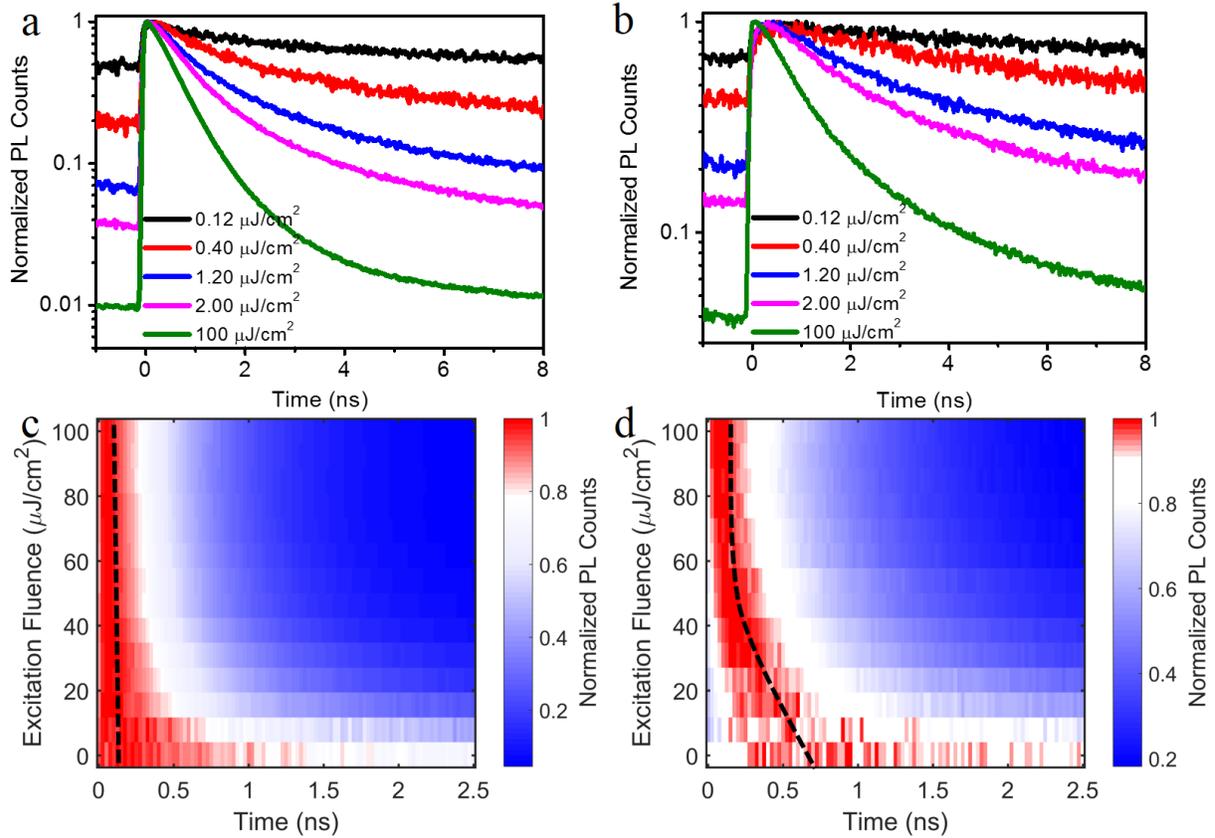

**Figure 2. Fluence-dependent TRPL of MAPbI$_3$ thin film at 77 K.** (a) The excitation fluence dependent TRPL of the HE peak A. (b) The excitation fluence dependent TRPL of the LE peak B. (c) and (d) are color plots of the detailed fluence-dependent TRPL study of the HE and LE peak, respectively. The black dashed lines are the eye-guide for the evolution of the HE peak and LE peak TRPL maxima positions as the excitation fluence increases.

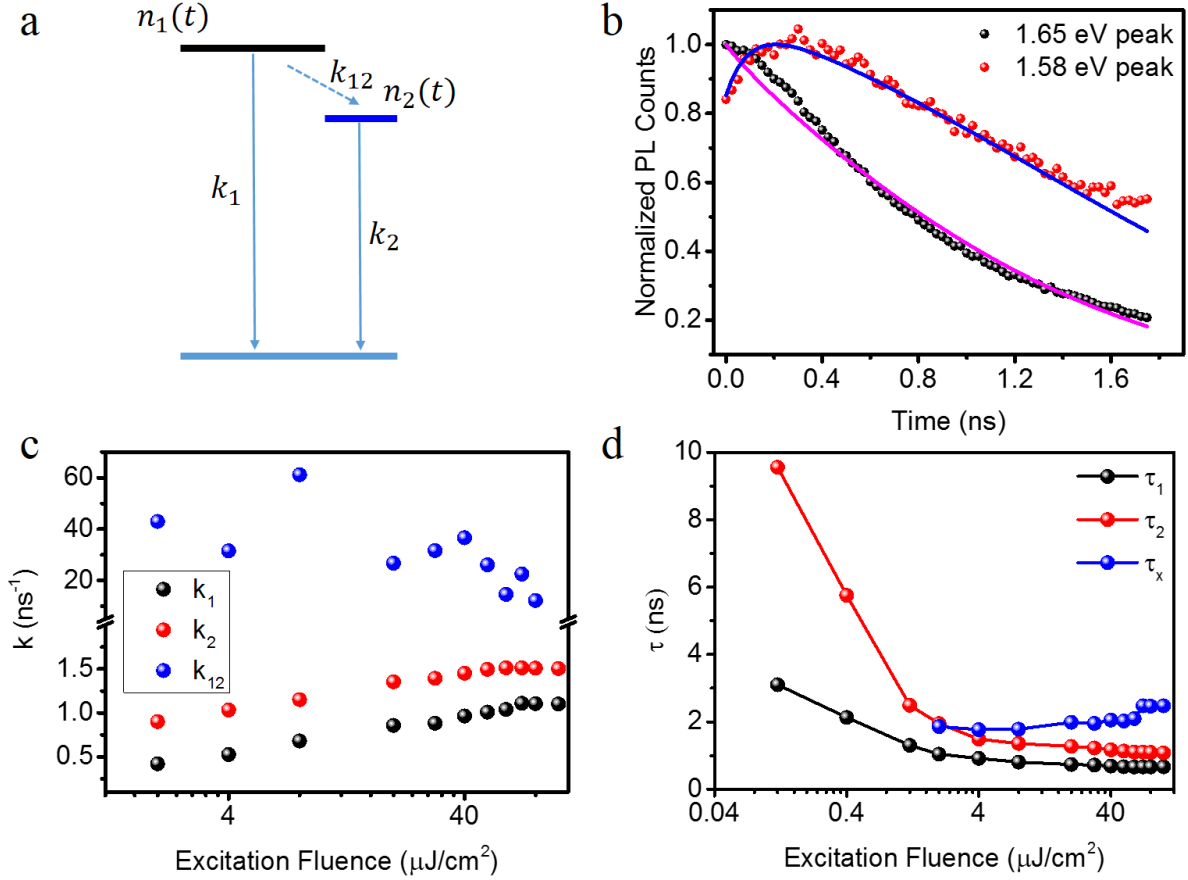

**Figure 3. Excitation fluence dependent kinetic rates in MAPbI3 thin film at 77 K.** (a) Schematic of the two-state recombination mechanism. The carrier density at the HE state and LE state are denoted as $n_1(t)$ and $n_2(t)$, respectively. $k_1$ and $k_2$ are the direct decay rate of HE state and LE state, respectively. And $k_{12}$ is the rate constant of injection from the HE state to the LE state. (b) Normalized TRPL at the excitation fluence of 2.0 μJ/cm², and the solid lines are the fitting results based on the proposed model shown in (a). (c) Fluence-dependent rate constants obtained by fitting the experimental TRPL data in Fig. 2 using the proposed model shown in (a). (d) Fluence-dependent lifetime of the HE state ($\tau_1$) (black dots) and LE state ($\tau_2$) (red dots) obtained by a mono-exponential fitting. $\tau_x$ (blue dots) is the temporally averaged injection time.

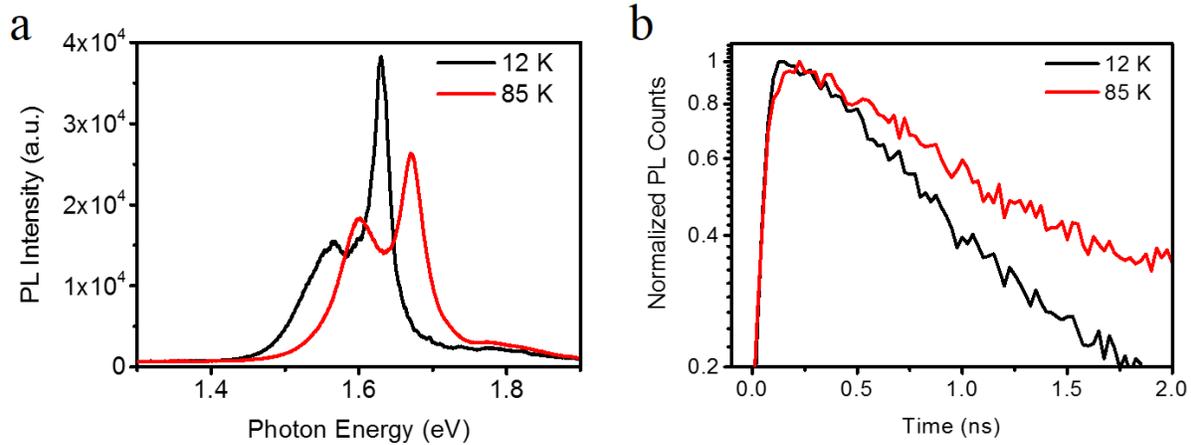

**Figure 4. Temperature dependence of the PL spectra.** (a) PL spectra at 12 K (black) and 85 K (red) excited by 532 nm pulsed laser (filtered from the supercontinuum laser) with the fluence of 12 μJ/cm$^2$. (b) Normalized TRPL of the LE state at 12 K (black) and 85 K (red).

**Supporting Information**

Supporting Information is available from the Wiley Online Library or from the authors.

**Acknowledgment**

This work was supported by the NY State Empire State Development's Division of Science, Technology and Innovation (NYSTAR) through Focus Center-NY–RPI Contract C150117. The optical characterization was supported by Rensselaer Polytechnic Institute (RPI) and the Center for Future Energy Systems (CFES), a New York State Center for Advanced Technology at RPI. T. Ye and H. Li were supported by the National Natural Science Foundation of China (51625304). The perovskite film fabrication was supported by Micro and Nanofabrication Clean Room (MNCR). Su-Fei Shi acknowledges support from the AFOSR through Grant FA9550-18-1-0312.

**Author Contribution**

S.-F. Shi conceived the experiment. Z. Li and H. Li fabricated the perovskite film. T. Wang, Y. Chen, Z. Li and Z. Lian performed the measurements. S.-F. Shi, Y. Chen, T. Wang and Z. Li analyzed the data. S.-F. Shi supervised the project. S.-F. Shi wrote the manuscript with the input from all the other co-authors. All authors discussed the results and contributed to the manuscript.

**Competing interests**

The authors declare that they have no competing interests.